\documentclass[fleqn,10pt]{wlscirep}
\usepackage{amssymb,amsmath}
\usepackage{color}

\newcommand{\blue}{\textcolor{black}}
\title{Orientational Mapping Augmented Sub-Wavelength Hyper-Spectral Imaging of Silk}
\author[1]{Meguya Ryu}
\author[2,3]{Armandas Bal\v{c}ytis}
\author[2]{Xuewen Wang}
\author[4]{Jitraporn Vongsvivut}
\author[5]{\\\protect Yuta Hikima}
\author[6]{Jingliang Li}
\author[4]{Mark  J. Tobin}
\author[2,7,8*]{Saulius Juodkazis}
\author[4,*]{\\\protect Junko Morikawa}
\affil[1]{Tokyo Institute of Technology, Meguro-ku, Tokyo
152-8550, Japan} \affil[2]{Nanotechnology facility, Center for
Micro-Photonics, Swinburne University of Technology, John st.,
Hawthorn, Victoria 3122, Australia}\affil[3]{Department of Laser
Technologies, Center for Physical Sciences and Technology,
Savanoriu Ave. 231, LT-02300 Vilnius, Lithuania}\affil[4]{Infrared
Microspectroscopy Beamline, Australian Synchrotron, Clayton,
Victoria 3168, Australia}\affil[5]{Department of Chemical
Engineering, Graduate School of Engineering, Kyoto University,
Nishikyo-ku, Kyoto 615-8510, Japan}\affil[6]{Institute for
Frontier Materials, Deakin University, Waurn Ponds, Victoria 3217,
Australia}\affil[7]{Center of Nanotechnology, King Abdulaziz
University, Jeddah 21589, Saudi Arabia}\affil[8]{Melbourne Center
for Nanofabrication, Australian National Fabrication Facility,
Clayton 3168, Australia} \affil[*]{SJ: sjuodkazis@swin.edu.au; JM:
morikawa.j.aa@m.titech.ac.jp} 

\begin{abstract}
Molecular alignment underpins optical, mechanical, and thermal
properties of materials, however, its direct measurement from
volumes with micrometer dimensions is not accessible, especially,
for structurally complex bio-materials. How the molecular
alignment is linked to extraordinary properties of silk and its
amorphous-crystalline composition has \blue{to be} accessed
\blue{by a} direct measurement from a single silk fiber. Here, we
show orientation mapping of the internal silk fiber structure via
polarisation-dependent IR absorbance at high spatial resolution of
\blue{$4.2~\mu$M and} $1.9~\mu$M in a hyper-spectral IR imaging by
attenuated total reflection using synchrotron radiation in the
spectral fingerprint region around $6~\mu$M wavelength.
Free-standing longitudinal micro-slices of silk fibers, thinner
than the fiber cross section, were prepared by microtome for the
four polarisation method to directly measure the orientational
sensitivity of absorbance in the molecular fingerprint spectral
window of the amide bands of $\beta$-sheets and amorphous
polypeptides of silk. \blue{Flat lateral micro-slices of silk
eliminates shape related artefact in determination of absorbance
anisotropy and order parameters of the amide bands.}
\end{abstract}
\begin{document}

\flushbottom \maketitle

\thispagestyle{empty}
\section*{Introduction}

The infra-red (IR) spectral region from 3-10~$\mu$m, referred to
as the fingerprint region, is used for the quantitative analysis
of molecular species in a wide range of applications spanning
fields of climate change~\cite{Yang}, environmental
monitoring~\cite{Lu}, bio-medical~\cite{Stanley}, material
science~\cite{Ulrike}, and security~\cite{Schliesser}. All imaging
methods have mounting challenges to characterise volumes with
cross sections approaching the wavelength of the utilised light.
In the UV-visible and IR spectral domains, near-field techniques
using sharp nano-tips and plasmonic enhancement are used to reach
nanoscale spatial resolutions, usually at the expense of
polarisation information. However the application of polarised
light permits analysis of the molecular orientation and chirality,
which define mechanical, thermal, and optical
properties~\cite{Yoshioka}. At different wavelengths it is
possible to access orientational information of hierarchical
structures which underpins mechanical material properties which we
could harness by engineering their artificial
counterparts~\cite{Ulrike}.

Fourier transform IR (FT-IR) spectroscopy, when combined with a
microscope accessory, provides hyper-spectral imaging when
spectrally broadband or wavelength-tunable excitation sources are
utilised. In the IR spectral range, a combination of
sub-wavelength spatial resolution to characterise the anisotropy
of absorbance due to local molecular orientation and spatial 2D
(3D) mapping would enhance current analytical techniques and has
high potential in material and bio-medical fields. In addition the
use of a  synchrotron  beam  offers  highly  collimated IR
radiation with $10^2 - 10^3$ times higher brightness than that
available from laboratory-based IR sources (Globar$^\circledR$).
Such a unique characteristic enables the acquisition of
high-quality FT-IR spectra at diffraction-limited spatial
resolution, making synchrotron-IR microspectrscopy an excellent
analytical platform for acquiring spatially resolved chemical
images of materials at a lateral resolution between 3-10~$\mu$m.
Using attenuated total reflection (ATR) with a high refractive
index $n = 4$ Ge contact lens, a state-of-the-art resolution of
$1.9~\mu$m, which is sub-wavelength in the IR molecular finger
printing spectral range, can be achieved and was one of the aims
of this study.

The field of bio-medical applications could be one of the main
beneficiaries of high-spatial resolution techniques with a focus
on sensors and bio-materials. In protein based materials, the
molecular ordering, orientation, and conformation define their
properties~\cite{Yoshioka}. Silk was the material of choice in
this study due to its bio-compatibility and bio-
degradability~\cite{Tao,Li}. It has high mechanical strength with
rich structural and compositional complexity ranging from
$\alpha$-coils (IR absorbance at 1660~cm$^{-1}$), metastable
$\beta$-turns (silk I), crystalline $\beta$-sheets (silk II), and
amorphous random fibroin  protein structure~\cite{Liu}. Controlled
modification of silk structure from  water soluble amorphous phase
to crystalline $\beta$-sheets is a current focus of
research~\cite{Qin,Schick,16b054101}, with structural
characterisation of silk having been carried out with X-ray
diffraction (XRD), nuclear magnetic resonance (NMR), and IR
spectroscopy of silk fiber bundles and amorphous
powders~\cite{Asakura,17m356}.

\blue{A} systematic study on orientational properties of the
building blocks of the crystalline-amorphous hierarchial structure
of single silk fibers which  is essential to understanding the
properties of silk, e.g., why a faster reeling makes stronger
fibers~\cite{Shao,Du} and how it is linked to fragility and
relaxation in polymers~\cite{Dalle}. \blue{Structure of single
spider silk fibers was investigated by XRD including changes due
to water uptake~\cite{Riekel,Sampath}. Differences of spectral
band positions using free space IR and ATR-IR
spectroscopies~\cite{BouletAudet,BouletAudet2} and order parameter
determination~\cite{Paquet,Cruz} have been carried out for single
fibers. Synchrotron X-ray microscopy was used to reveal
orientational effects in absorbance of spider silk at high spatial
resolution $\sim 50$~nm averaged over the entire fiber
thickness~\cite{Cruz,Rousseau1}.} A polarisation dependence of the
IR absorbance of amides in silk fiber can provide deeper insights
in molecular orientation of hierarchial silk structure, which for
example is known to define thermal conductivity, $\kappa$, which
is increasing in the stretched form
$\kappa\sim\sqrt{E}$~\cite{Huang} ($E$ is the Young's modulus) and
is increasing under strain towards the onset of melting at around
$200^\circ$C~\cite{Martel}. In the presence of hydrogen bonding,
the orientation is linked to an increased
crystallinity~\cite{Zhang,Zhang1}. Nanoscale orientation of
proteins and their 3D conformation are at the core of their
optical, mechanical, thermal, and bio-functions. \blue{These}
important properties \blue{can be better understood using} high
resolution \blue{techniques, which have to be applied}
simultaneously \blue{for} space and spectrum \blue{measurements to
unveil primary and secondary} molecular orientation/alignment.
\blue{The} polarisation dependence of the absorbance bands
\blue{is used to determine anisotropy of absorbance in silk. It
allows to investigate structure of silk at nanoscale~\cite{Paquet}
and relate it to the hierarchical structure and mechanical
properties~\cite{Papadopoulos,Papadopoulos1,Ene}. }

Here, sub-wavelength spatial resolution was combined with
hyper-spectral imaging to characterise local absorbance of silk
fibers modified by ultra-short laser pulses using the in-house
developed ATR FT-IR instrument at Australian Synchrotron.
Polarisation dependence of the absorbance was successfully invoked
to reveal the high degree of orientation of amide building blocks
of silk in fibers and \blue{to recognise} laser-induced
amorphisation. \blue{In order to exclude shape related effects in
absorbance measurements and to reveal molecular orientation along
the silk fiber, thin and flat microtome slices of lateral silk
fiber cross sections were prepared and used in this study.}

\begin{figure}[tb]
\begin{center}
\includegraphics[width=8.50cm]{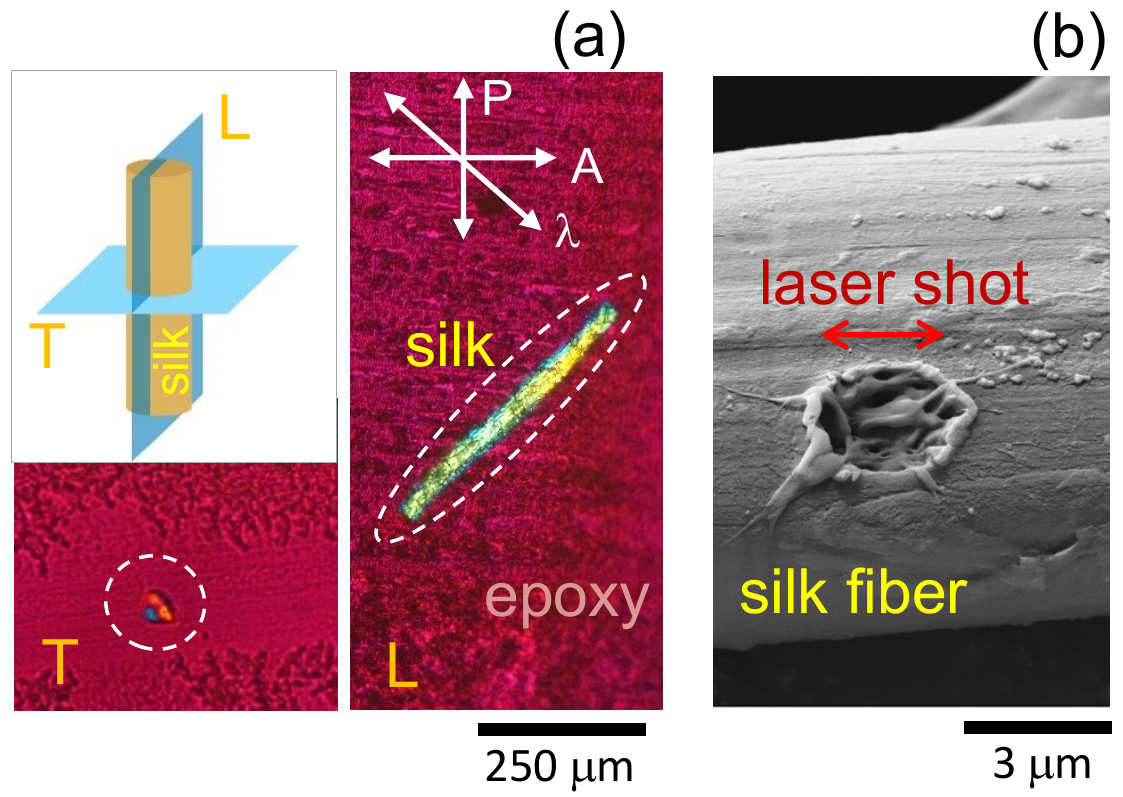}
\caption{(a) Longitudinal (L) and transverse (T) 5-10~$\mu$m-thin
slices of silk embedded in epoxy were used for FT-IR
micro-spectroscopic characterisation. Cross-polarised optical
image of the sliced silk fiber obtained with a waveplate
($\lambda$) shifter. Sample preparation: silk fiber was aligned
and epoxy embedded, microtome sliced for L and T directions. (b)
SEM image of a single laser pulse melted silk; laser wavelength
512~nm, pulse duration 230~fs, focused with objective lens with
numerical aperture $NA = 0.5$, pulse energy 85~nJ, linear
polarisation was along the fiber (marked by arrow).}
\label{f-sample}
\end{center}
\end{figure}

\begin{figure}[tb]
\begin{center}
\includegraphics[width=7.50cm]{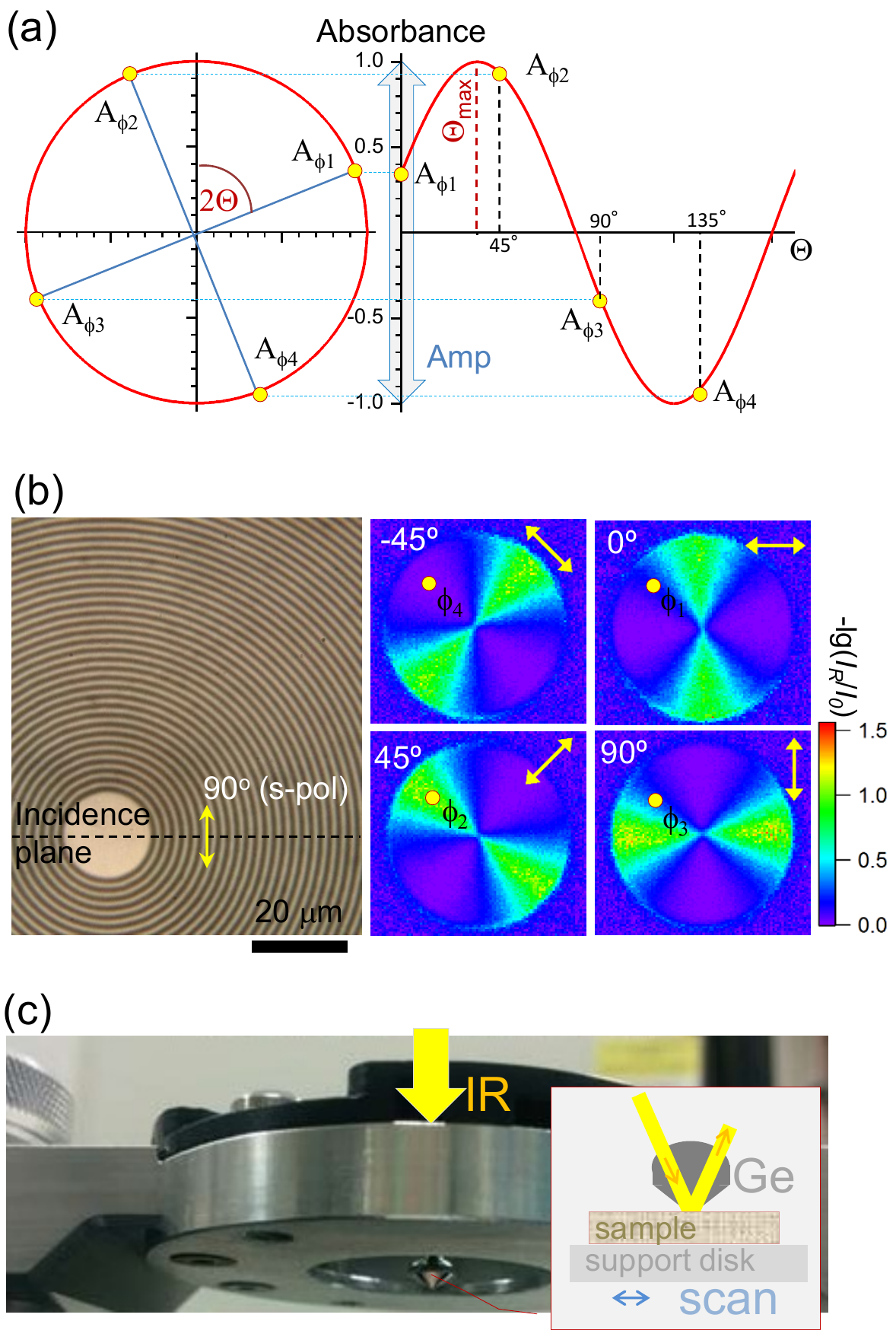}
\caption{(a) Four-polarisation method in IR absorbance at
different polarisation azimuths, $A=f(\Theta)$,~\cite{Hikima},
which was used to determine the orientational anisotropy in
far-field FT-IR measurements. (b) An optical image of a concentric
2-$\mu$m-period Au grating  (with a duty  cycle of  0.5) on a
cover glass used for reflection FT-IR imaging carried out with a
Cassegrainian objective lens with a focusing cone over a
17-to-30$^\circ$ range; dashed-line marks the top-view of the
incidence plane. The diameter of the circular grating was 0.5~mm.
Four reflection maps for a linearly polarised incidence at the
specified polarisation angle (schematically also marked by arrows)
at 1500~cm$^{-1}$ ($\sim 6.7~\mu$m) wavelength (Spotlight,
PerkinElmer); $I_{R,0}$ are the reflected and incidence
intensities, respectively. (c) In-house developed ATR accessory
for Hyperion 2000, Bruker microscope. Inset shows schematic
illustration of optical geometry used in the ATR FT-IR measurement
with a 100-$\mu$m diameter Ge tip (refractive index $n = 4$). }
\label{f-method}
\end{center}
\end{figure}

\section*{Experimental}

Silk samples were cut to a thickness of a few micrometers by
microtome (Fig.~\ref{f-sample}), then laser modified by single
laser shots before FT-IR measurements at the IR Microspectroscopy
Beamline (Australian Synchrotron) using a polarisation
discrimination method for the far-field absorbance
measurement~\cite{Hikima} and subsequently at a high spatial
resolution using in-house developed ATR accessory
(Fig.~\ref{f-method}).

\subsection*{Silk micro-slices}

Domestic silk (\emph{Bombyx mori}) fibers were used for
experiments after removal of sericin rich
cladding~\cite{16b054101}. For the cross-sectional observation,
the natural silk fibers were aligned and embedded into an epoxy
adhesive (jER 828, Mitsubishi Chemical Co., Ltd.). Fibers fixed in
the epoxy matrix were cut in 1-5~$\mu$m-thick slices which were
found to possess sufficient mechanical robustness for the FT-IR
transmission measurements carried out without any supporting
substrate.  This was important to increase sensitivity of the
far-field absorbance measurements and to decrease reflective
losses that may occur through use of a supporting substrate.
Longitudinal (L) and transverse (T) slicing of the silk fibers was
carried out by microtome (RV-240, Yamato Khoki Industrial Co.,
Ltd.; see Fig.~\ref{f-sample}). The slices, which were cut thinner
than the original silk fibers, were used for the transmission
measurements in mapping mode along and across the fiber without
background interference from a supporting epoxy host. For the ATR
FT-IR, an aluminium disk was used to mount the thin fiber cross
section, which was subsequently brought into contact with a
100-$\mu$m-diameter sensing facet of the Ge ATR hemisphere
(refractive index $n = 4$).

Modification of silk was carried out using 515~nm wavelength and
230~fs duration pulses (Pharos, Light Converison Ltd.) focused
with an objective lens of numerical aperture $NA = 0.5$
(Mitutoyo). Single pulse modifcations were carried out with pulse
energy, $E_p$, indicated at the irradiation point, using an
integrated industrial laser fabrication setup (Workshop of
Photonics, Ltd.). Optical and scanning electron microscopy (SEM)
were used for structural characterisation of the laser modified
regions.

\begin{figure}[tb]
\begin{center}
\includegraphics[width=8.50cm]{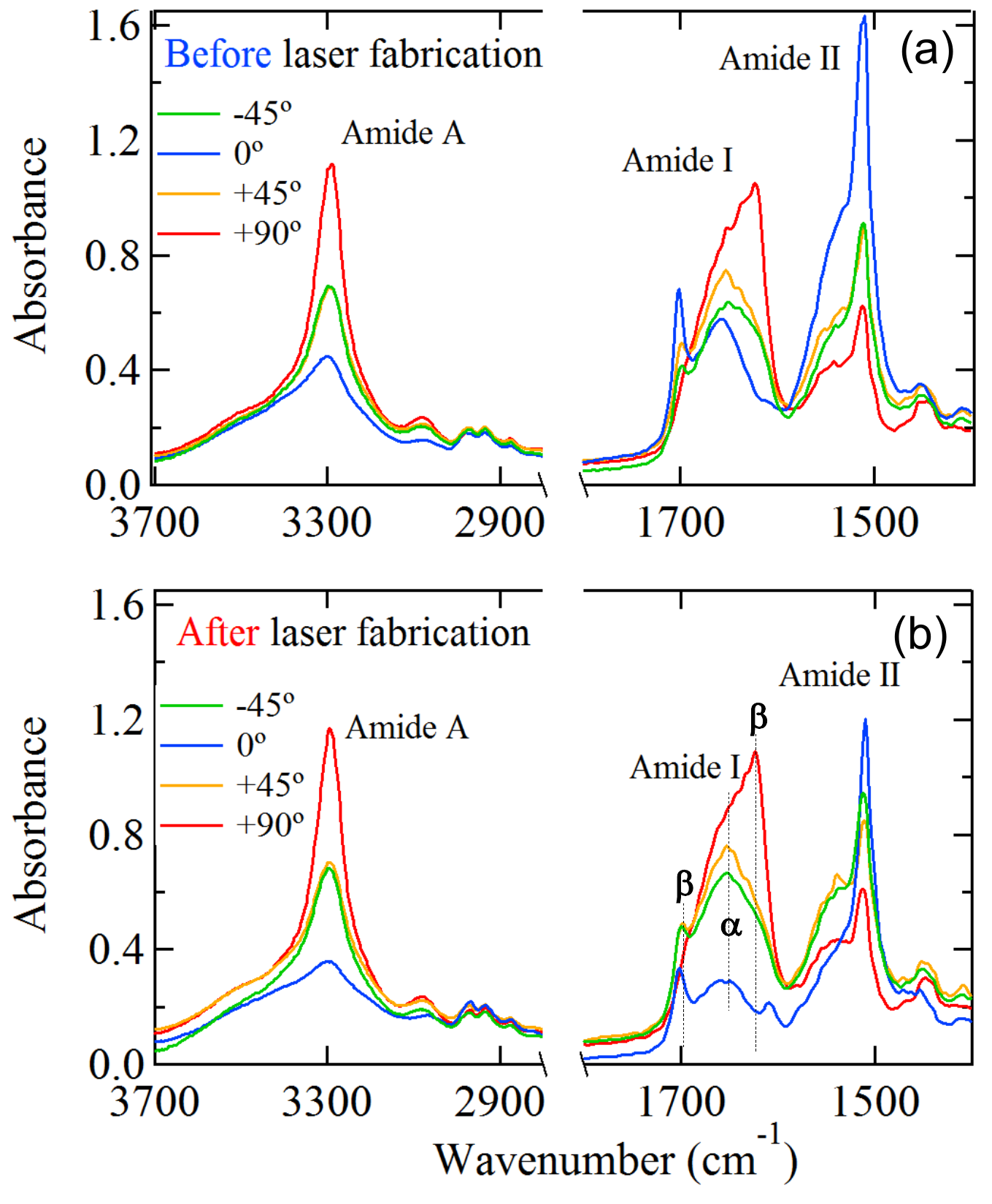}
\caption{Polarisation discriminated absorbance spectra of pristine
(a) and laser 515~nm/230~fs irradiated (b) silk fiber; laser pulse
energy was 8.5~nJ and pulse-to-pulse separation of 2~$\mu$m in
xy-array. Area of laser patterning was $10\times 10~\mu$m$^2$; IR
beam diameter at focus on the sample was $4.2~\mu$m. }
\label{f-far}
\end{center}
\end{figure}

\begin{figure}[tb]
\begin{center}
\includegraphics[width=10.50cm]{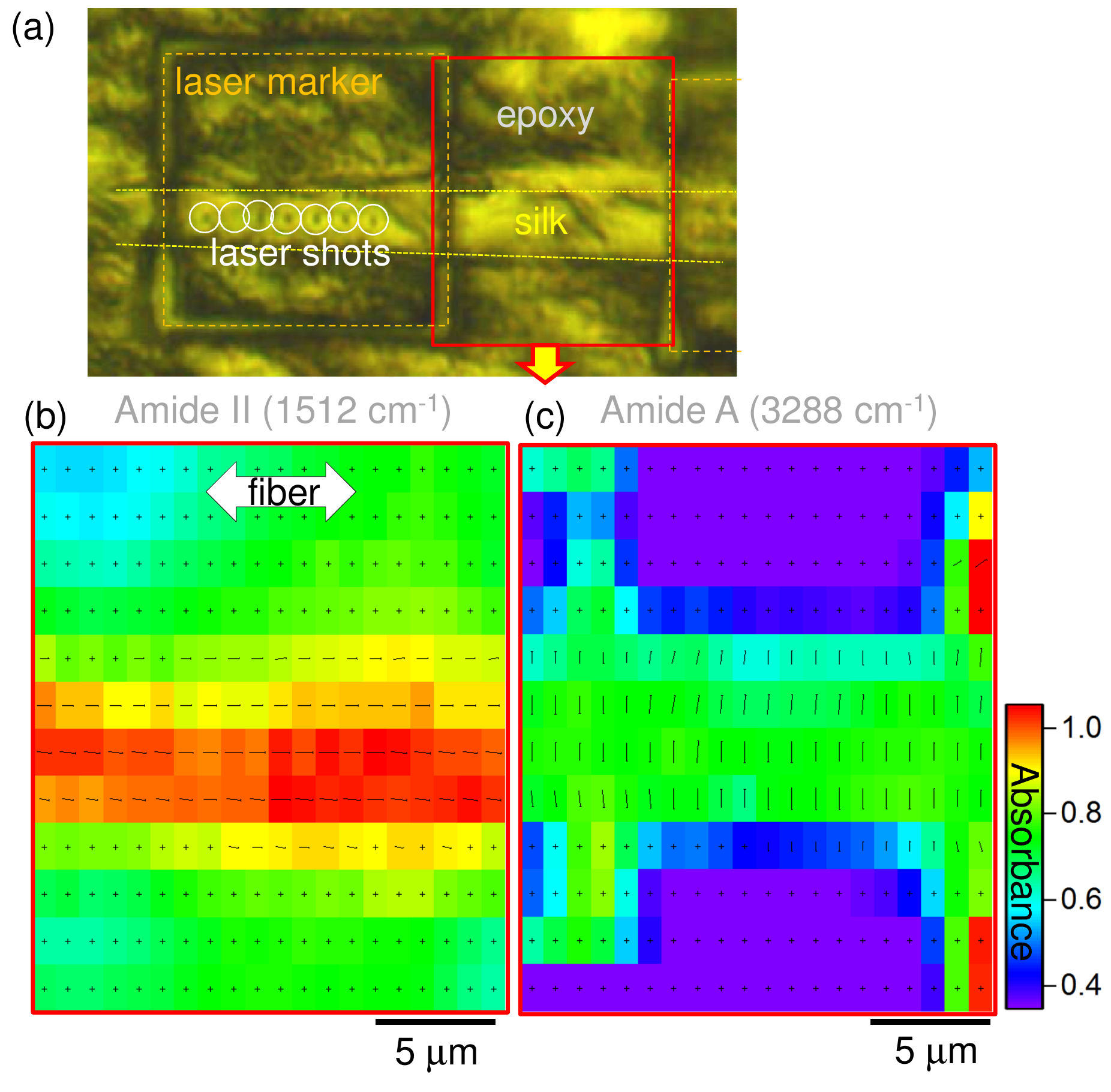}
\caption{(a) An optical image of the L cross section of a
\emph{Bombyx mori} silk fiber with laser marked $20\times
20~\mu$m$^2$  regions and laser irradiated spots.  The region
mapped in (b,c) is shown in a solid rectangle in (a). (b,c)
Orientation vector map (marker's length Eqn.~\ref{e1} and
orientation Eqn.~\ref{e2}) overlayed with the far-field absorbance
(color map) at the C-N (b) and N-H (c) bands; these bonds are
known to be perpendicular. \blue{S-polarised incident light was
perpendicular to the fiber; in the plane of image.}} \label{f-map}
\end{center}
\end{figure}

\subsection*{Four-polarisation method}

Anisotropy of the far-field absorbance can be quantified using the
four polarisation  method~\cite{Hikima} by  measuring absorbance
at four polarisations separated by a $\pi/4$ azimuth and assuming
a linear absorption of molecular dipoles in the E-field of light.
A sine wave profile of absorbance fit is expected
(Fig.~\ref{f-method}(a)) with the min-max amplitude of absorbance,
$Amp$, and dipole orientation angle, $\theta$ defined for each
pixel of a hyper-spectral image~\cite{Hikima}:
\begin{equation}\label{e1}
Amp = \sqrt{(A_{\phi_4} - A_{\phi_2})^2 + (A_{\phi_3} -
A_{\phi_1})^2},
\end{equation}
\begin{equation}\label{e2}
\theta = \frac{1}{2}\tan^{-1}\left(\frac{A_{\phi_3} -
A_{\phi_1}}{A_{\phi_4} - A_{\phi_2}}\right),
\end{equation}
\noindent where $A_{\phi_{1,2,3,4}}$ are absorbance at the four
polarisation azimuths separated by $\pi/4$; $Amp =
A_{max}-A_{min}$ is defined by the maximum and minimum
absorbances.

This four-polarisation method was implemented using a
Cassegrainian FT-IR objective with the linear polarisation set
right at the entrance of the objective lens by a wire-grid
polariser. To test the validity of the four-polarisation method
for this geometry, where two reflections on curved mirrors are
encountered by linearly polarised incident beam in the
Cassegrainian optic, a circular grating reference sample was made
by electron beam lithography (EBL; ACE-7000/EBU, Sanyu Electron
Ltd.) and standard lift-off method. A 30nm-thick Au coating was
thermally evaporated on a 10~nm adhesion layer of Cr on a cover
glass for the lift-off  over  the  EBL  defined  circular  pattern
in ZEP520 resist; diameter of the circular grating was 0.5~mm. The
grating with a width of Au rings of 1~$\mu$m and period of
2~$\mu$m represents a reflective sub-wavelength pattern  of  a
constantly changing  orientation  at  the  IR wavelength of
1500~cm$^{-1}$  or $\sim 6.7~\mu$m (Fig.~\ref{f-method}(b)).  By
setting four polarisations with a $\pi/4$ separation at incidence,
the reflection maps from the circular grating measured with
Spotlight, PerkinElmer are shown in Fig.~\ref{f-method}(b).
Angular integration of the reflected intensity at any radial
position closely followed the postulated sine wave rule (Fig.
2(a)); e.g., the four selected angle positions on the reflection
maps are marked by $\phi_{1,2,3,4}$ and follow intensity changes
by the sine wave form. The strongest reflection was observed for
the polarisation which is tangential to the circumference of the
grating ring  pattern.

\subsection*{High spatial resolution FT-IR spectroscopy}

The far-field transmission measurements were carried out with a
$NA = 0.5$ and $36^\times$ magnification Cassegrainian objective
lens. A wire-grid ZnSe polariser was used to set linear
polarisation \blue{(Specac Ltd., Kent, UK)}.

Synchrontron IR microspectroscopic measurement was performed using
a Bruker  Hyperion 2000 FT-IR microscope (Bruker Optik GmbH,
Ettlingen, Germany) coupled to a Vertex  V80v  FT-IR spectrometer,
and equipped with a liquid nitrogen-cooled narrow-band mercury
cadmium telluride (MCT) detector. As illustrated in
Fig.~\ref{f-method}(c), the in-house developed ATR FT-IR accessory
equipped with a 100-$\mu$m-diameter facet Ge ATR crystal was used
to acquire chemical images of the silk cross sections at a high
speed and a spatial resolution down to 1.9~$\mu$m~\cite{Pimm}. The
Ge contact lens of $NA = n\sin\varphi \simeq 2.4$ was used with $n
= 4$ and the $\varphi = 36.9^\circ$ half-angle of the focusing
cone. Deep sub-wavelength resolution $r = 0.61\lambda_{IR}/NA
\simeq 1.5~\mu$m is achievable for the IR wavelengths of interest
at the amide band of $\lambda_{IR} = 1600-1700$~cm$^{-1}$ or 6.25
- 5.9~$\mu$m. Use of the solid immersion lens also leads to a
reduction of the mapping step size by $\sim$4 times relative to
the stage step motion and was 250~nm. The far-field transmission
measurements were carried out with a $NA = 0.5$ and $36^\times$
magnification Cassegrainian objective. A wire-grid polariser was
used to set linear polarisation.

\section*{Results}

\subsection*{Polarisation dependence at single point}

Figure~\ref{f-far} shows absorbance of silk measured in
transmission for four different azimuthal orientations of the
linear polarisation with an angular separation of $\pi/4$ for silk
(\emph{Bombyx mori}) from laser exposed (a) and un- treated (b)
regions. A xy-array of laser irradiated spots at 8.5~nJ/pulse was
patterned with $2~\mu$m period while the IR spectra were acquired
from a 4.2$~\mu$m spot. The Amide I and II
regions~\cite{Taddei2005} were investigated for structural and
compositional changes induced by laser irradiation. The Amide II
band at 1508~cm$^{-1}$ is assigned to $\beta$-sheet secondary
structure, whilst the peak at 1546~cm$^{-1}$ is associated with
disordered (amorphous) fibroin. The Amide I band follows a similar
distribution with components at $\sim$1625~cm$^{-1}$
($\beta$-sheets) and 1648~cm$^{-1}$, which are associated with
irregular structures including random coil and extended
chains~\cite{Taddei2005}. Other characteristic bands are
associated with Silk I, type II $\beta$-turns
(1647--1654~cm$^{-1}$), $\alpha$-coils (1658--1664~cm$^{-1}$) as
well as turns and bends 1699~cm$^{-1}$~\cite{QLu2010}.

\begin{figure}[tb]
\begin{center}
\includegraphics[width=7.0cm]{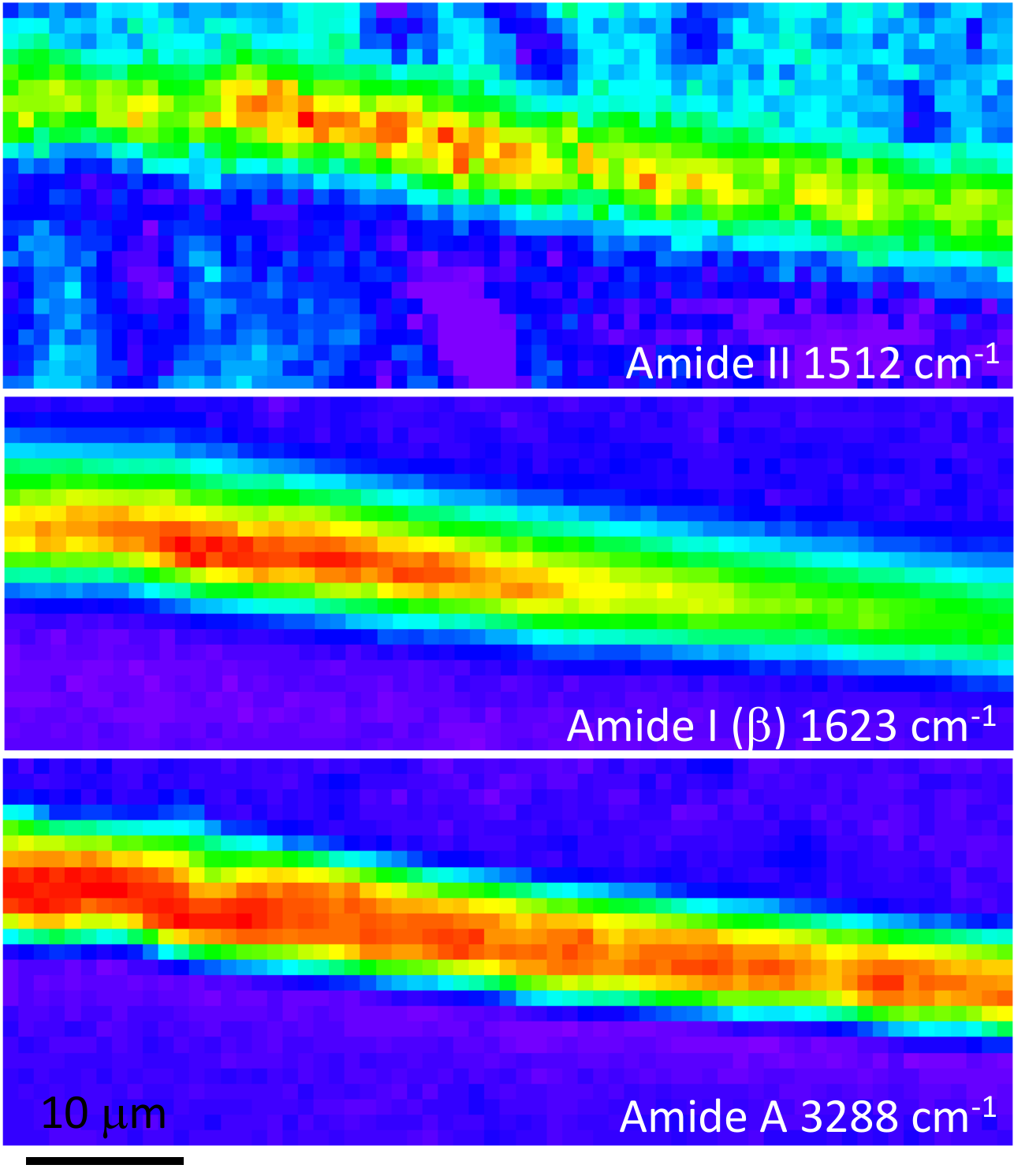}
\caption{High resolution $1.9~\mu$m ATR FT-IR maps at $1.9~\mu$m
resolution of the longitudinal (L) cross sections of silk
presented in auto-scale for better viewing; a background-corrected
absorbance is ranging from 0 to approximately 0.2. Lateral step
size between pixels was 0.5~$\mu$m; as-measured pixelated
absorbance maps are presented. \blue{Polarisation of incident
light onto ATR prism was $s$ (in the plane of image; along
y-axis).}} \label{f-L}
\end{center}
\end{figure}

\begin{figure}[tb]
\begin{center}
\includegraphics[width=7.50cm]{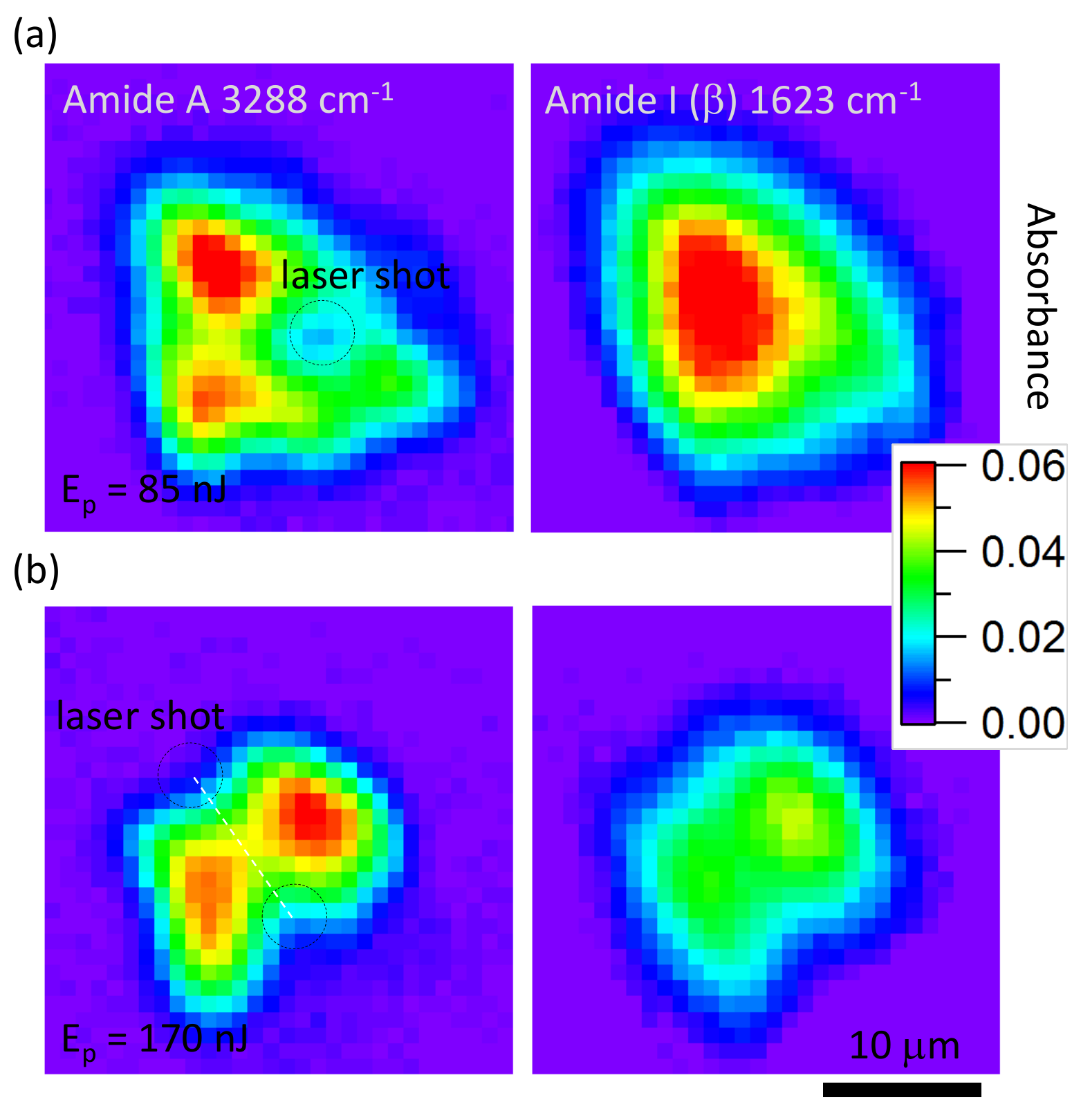}
\caption{High resolution $1.9~\mu$m ATR maps of transverse (T)
cross sections of silk (\emph{Bombyx mori}) with laser
515~nm/230~fs irradiated spots; laser pulse energies, $E_p =
85$~nJ (a) 170~nJ (b) on the sample; the polarisation was linear.
The lateral step size was 0.5~$\mu$m; as-measured pixelated
absorbance maps are presented. \blue{Polarisation of incident
light onto ATR prism was $s$ (in the plane of image; along
y-axis).}} \label{f-laser}
\end{center}
\end{figure}

Laser irradiation was found to strongly affect the sharp
absorbance peak at 1700~cm$^{-1}$ when laser pulse energy was
exceeding the threshold of structural damage at $E_p \simeq 8$~nJ
for the used focusing (Fig.~\ref{f-far}). This is indicative of
amorphisation, which would be expected based on the observed
changes in the SEM images of the laser exposed silk shown in
Fig.~\ref{f-sample}(b). A distinct polarisation dependence was
also observed, as expected from a crystalline rich ($\sim 85\%$)
silk fibers at the $\beta$-sheet Amide I band. The strongest
absorbance at the Amide II (C-N) \blue{stretching} band at
$0^\circ$ polarisation corresponded to the C-N bond, which is
aligned along the fiber direction. The Amide \blue{I} (C=O)
\blue{stretching} band, on the other hand, showed an inverse
correlation with the polarised absorbance spectrum of the Amide II
band which was strongest at the perpendicular polarisation
(Fig.~\ref{f-far}) as expected from previous Raman scattering
studies~\cite{Rousseau}. The N-H \blue{stretching} band showed the
same polarisation dependence as the C=O \blue{stretching} band.
Due to such a unique and strong polarisation dependence of the
absorbance at a single spot (Fig.~\ref{f-far}), the far-field
transmission measurement in the mapping mode was subsequently
performed to gain additional insights into molecular
orientation/alignment along the length of silk fibers made
accessible via microtomed L cross sections.

\subsection*{Molecular orientation in silk: far-field case}

The four polarisations method was applied to reveal orientational
association  of the amide bands using Eqns.~\ref{e1} and \ref{e2}.
Figure~\ref{f-map} shows the chemical maps of the L cross section
of silk fiber with measured $\sim 4.2~\mu$m spatial resolution
($NA = 0.5$). Mapping data (as measured) are visualised by
overlaying absorbance at the selected wavenumber values of Amide
II and Amide A at 1512~cm$^{-1}$ and 3288~cm$^{-1}$, respectively.
The corresponding vector plot (markers' length Eqn.~\ref{e1} and
orientation Eqn.~\ref{e2}) revealed that the orientation is
horizontal and the amplitude $Amp$ is proportional to the length
of the bar-marker ($\theta = 0^\circ$ is horizontal). The vector
plot represents a background-free component of absorbance change
caused by a change in molecular alignment. Perpendicular
orientation between C=O and C-N bonds observed in the single spot
spectrum (Fig.~\ref{f-far}) has been confirmed for the
non-irradiated silk regions (Fig.~\ref{f-map}(b) vs (c)).
Nevertheless, some of the Amide II bands present in the epoxy
matrix were found to possess a random orientation.

To quantify the order the standard second momentum $P_2(\theta)$
of the orientation function also known as the Herman's function
can be expressed via the absorbance ratio at two perpendicular
linear polarisations, the dichroic ratio, $D =
A_{0^\circ}/A_{90^\circ}$ (see Supplement for details)
as~\cite{Hikima,Hashimoto}:
\begin{equation}\label{e3}
P_2(\theta) = \frac{3\langle\cos^2\theta\rangle - 1}{2},
\end{equation}
\noindent where $\theta$ is the angle between the transition
dipole moment and the selected orientation (along silk
fiber). 
The second momentum of the C=O (Amide I) band was found
$P_2(\theta)=-0.36$ from the Raman measurements~\cite{Rousseau}
(-0.5 corresponds to a pure perpendicular orientation to the fiber
axis). A slightly less ordered C=O bonds were determined in this
study with \blue{$P_2(\theta) = -0.29\pm 0.026$} (see, Supplement
for details). The difference can be accounted by a fiber caused
anisotropic diffraction in the case of Raman measurements while a
flat cross section was used in this study. The order analysis
reveals that silk fibers are up to $\sim 60\%$ crystalline (see,
Supplement) which is approximately twice larger than observed by
synchrotron FT-IR in regenerated silk fibroin after
crystallisation in alcohol bath $\sim 28\%$~\cite{Ling}.

\subsection*{High resolution ATR mapping}

The highest spatial resolution was achieved using ATR FT-IR,with a
$NA = 2.6$ focus, realised using the combination of a germanium
solid immersion lens of $n = 4$, with a Cassegrainian objective of
$NA = 0.6$ and $20^\times$ magnifcation. Although no polariser was
used for the mapping, synchrotron beam had a dominant s-polarised
linear component. It should be emphasised, that  in addition to
the enhanced lateral spatial resolution this ATR FT-IR optical
configuration offered high surface sensitivity due to a low
penetration depth of $\sim \lambda_{IR}/4 \sim 1.5~\mu$m of the IR
radiation at the amide I absorption peak. Figure~\ref{f-L} shows
the highest spatial resolution $r \simeq 1.9~\mu$m chemical maps
of the L section of silk at a few spectral regions of interest
selected from a single hyper- spectral FT-IR data set. The Amide A
(N-H) band appeared to have the most uniform distribution across
the fiber compared to those of the Amide I (C=O) and $\beta$-
sheet, which were highly localised inside the core of the fiber.
This could be rationalised by the low sensitivity of N-H
absorption to a surrounding hierarchial structure of the protein
matrix, mainly, because of a low mass of hydrogen.

The distinct effect of laser irradiation on the molecular
structure of silk fibers was revealed in this study for the first
time by high-spatial resolution ATR FT-IR mapping of T-cross
sections (Fig.~\ref{f-laser}). Single laser shots created an
approximately 2~$\mu$m diameter ablation pits observable through
both optical or/and SEM images (Fig.~\ref{f-map}(a)). The pattern
of irradiation spots was controlled with a high precision of $\sim
50$~nm. This was instrumental in identifying the irradiation
locations on the absorbance maps (Fig.~\ref{f-laser}). Central
localisation of $\beta$-sheets is well distinct in the T sectional
images (Fig.~\ref{f-laser}).

\begin{figure}[tb]
\begin{center}
\includegraphics[width=7.50cm]{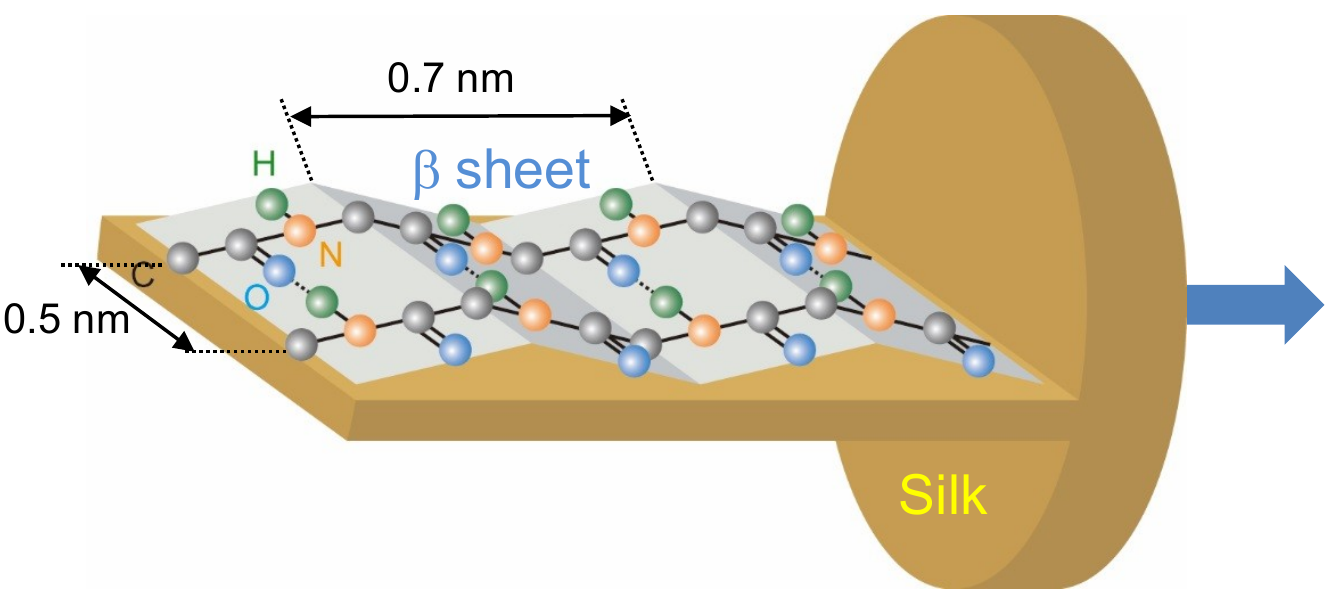}
\caption{The orientation of the C=O, C-N, and N-H bonds in amide
structure of the L-section of silk fiber~\cite{Alberts,Cruz}
\blue{confirmed in this study by the hyper-spectral imaging} (see,
Fig.~\ref{f-map}). Only the in-plane components of $\beta$-sheets
are drawn without out-off-plane alkyl moieties; hydrogen bonding
responsible for $\beta$-sheet crystallisation is shown by the
dotted line O$\cdot\cdot\cdot$H. An arrow marks fiber drawing
(strain) direction important to alignment of $\beta$-sheets.
\blue{Microtome slices allowed to measure absorbance of the
lateral flat cross sections without introduction of a fiber shape
related anisotropy. }} \label{f-stru}
\end{center}
\end{figure}

The lowest laser pulse energy which made recognisable
modifications to the fibres in a single shot was $E \simeq 8$~nJ.
Spectral maps in Fig.~\ref{f-laser}(a) show that only modification
at the Amide A band was observed after laser irradiation, while
Amide I $\beta$-sheet structure was not affected. By doubling the
laser pulse energy, the distribution of both the Amide A and I
bands were found altered (Fig.~\ref{f-laser}(b)).  This finding is
consistent with the chemical bond strength which are 189~kJ/mol
(Amide A) and 1076.5~kJ/mol (Amide I) bonding of $\beta$-sheet,
respectively~\cite{crc}.

\section*{Discussion}

Substrate-free absorbance measurements of silk fibers, with
lateral resolution defined by $NA = 0.5$ for the far-field
transmission and $NA = 2.4$ for the ATR FT-IR hyper-spectral
mapping, have shown consistency between spatial localisation of
the Amide I and II bands in the silk fiber. The fiber core is
$\beta$-sheet enriched, hence, crystalline, as revealed by L and T
cross sections of silk fibers. \blue{Flat microtome slices
eliminated fiber shape related optical distortions and allowed
measurements of order parameters of the amide bands (see, online
supplement). Such L-cross sections can be also beneficial for
determination of order parameters by Raman scattering. }

The four polarisation method was adopted in transmission mode for
the high-brightness synchrotron IR radiation and applied to the L
section of silk fibers to reveal unambiguously the orientational
structure of the amide bands as illustrated in Fig.~\ref{f-stru}.
\blue{The Amide A (N-H) and amide I (C=O) have slightly different
$P_2$ order parameters. It was confirmed that Amide A and Amide II
bands are perpendicular (see, online supplement).} The spatial
mapping functionality demonstrated in this study possesses a
capability to reveal silk amorphisation mechanisms activated by
application of tightly focused ultra-short laser pulses. This
distinct laser irradiation is required for a fast thermal
quenching in excess of $2\times 10^3$~K/s for solidification of
amorphous silk melts~\cite{Schick}. Understanding the mechanisms
of amorphous fibroin crystallisation induced by ultra-short laser
pulses at the ablation threshold of glass
substrate~\cite{16b054101} requires structural sensitivity at high
spatial resolution to confirm the role of electron avalanche in
the formation of crystalline $\beta$-sheets in direct laser
printing~\cite{16le16133}. The 3D laser printing of silk scaffolds
by a direct write approach has a strong potential for bio-medical
implants, e.g., a plasma laser deposition of crystalline
silk~\cite{Tsuboi} and $\beta$-sheet formation form amorphous
fibroin under 266~nm laser irradiation~\cite{Tsuboi1} have been
demonstrated.  By applying stretching to films of pure sericin,
which is amorphous in silk fiber cladding, a molecular orientation
can be imprinted~\cite{Teramoto}.  Cast-drying of volumetric silk
workpieces for a mechanical post-processing in orthopedic
applications has been recently demonstrated with a need to control
nano-/micro-structure for the required specific strength and
modulus (stiffness)~\cite{Hotz} which can also be controlled by
molecular alignment.

\section*{Conclusions }

High spatial resolution has been achieved in hyper-spectral
imaging ATR FT-IR imaging  as demonstrated by the $\sim 1.9~\mu$m
($NA = 2.4$) resolution chemical imaging of silk at $\lambda_{IR}
\simeq 6~\mu$m wavelengths. It is shown that the four polarisation
method can be effectively used to reveal a prevalent orientational
ordering using far-field IR absorbance mapping. In silk, a strong
correlation between orthogonal C=O and C-N bonds has been
confirmed. \blue{The order parameters of the amides was determined
using micro-thin flat longitudinal microtome slices. For the C=O,
order parameter $P_2(\theta) = -0.30\pm 0.04$ and is comparable
with values obtained by different
methods~\cite{Rousseau,Paquet,Cruz}. This four polarisation method
can be used to recognise laser induced amorphisation of silk which
is water soluble.}

Insights into the orientational structure of biomaterials
responsible for their optical, mechanical, and thermal properties
is critical for applications and design of new materials. Here a
direct absorbance measurement of orientation of the chemical
bonding in silk at a record high spatial resolution is reported
using synchrotron based ATR FT-IR microspectroscopy. This
technique has been shown to possess potential as a powerful
analytical tool for a wide range of applications capable of
establishing the link between micro-/nano-structures and specific
properties of biomaterials.

\subsection*{Acknowledgements}

\small{J.M. acknowledges a partial support by a JSPS KAKENHI Grant
No. 16K06768. We acknowledge the Swinburne's startup grant for
Nanotechnology facility and partial support via ARC Discovery
DP130101205 and DP170100131 grants. The synchrotron-IR experiments
were performed through the merit-based access program (Proposal
ID. M11119) for the provision of the synchrotron beamtime at the
Australian Synchrotron IRM Beamline. Window on Photonics R\&D,
Ltd. is acknowledged for joint development grant and laser
fabrication facility.}


\section*{Author contributions statement}

\small{S.J. and J.M. initiated synchrotron proposal, J.M. and M.
R. proposed molecular alignment measurements, A.B., M.R., X.W.W.,
J.M., S.J. carried out experiment at Melbourne synchrotron on the
beamline under supervision of J.V. and M.T., X.W.W. made laser
irradiation of silk, M.R. carried our microtome and spectral
analysis, Y.H. and J.M. developed the four polarisation method,
silk samples were prepared in J.L. team. All the authors
participated in discussion and analysis of the results and
contributed to editing of the manuscript.}

\section*{Additional information}

\textbf{Competing financial interests} The authors declare no
competing financial interests.

\clearpage
\subsection*{Online Supplement}\label{online}
The order parameters and percentage of crystalline phase
($\beta$-sheets) in silk determined by different methods strongly
varies~\cite{Ling}. To quantify and visualise the order of
molecular alignment inside a T cross section of silk and to
compare with another synchrotron FT-IR absorbance
measurement~\cite{Ling}, the second order momentum $P_2(\theta)$
was calculated (Eqn.~2) for comparison with the orientation angle
$\theta$ (Fig.~4):
\begin{equation*}
P_2(\theta) = \frac{3\langle\cos^2\theta\rangle - 1}{2},
\end{equation*}
By using two IR absorptions at perpendicular and parallel
polarisations in respect to the selected orientation (along silk
fiber), the second order parameter $P_2$ can be written as
follows:
\begin{equation}\label{e22}
P_2(\theta) = \frac{A_1-A_2}{A_1+2A_2}
\end{equation}
where $\mathbf{n}$ is the direction of selected axis and $\theta$
is the angle of the transition dipole moment (Fig.~8). Incident
light polarization direction of $A_1$ is parallel while $A_2$ is
perpendicular to $\mathbf{X}_1$ axis.
\begin{figure}[htb]
\begin{center}
\includegraphics[width=4.50cm]{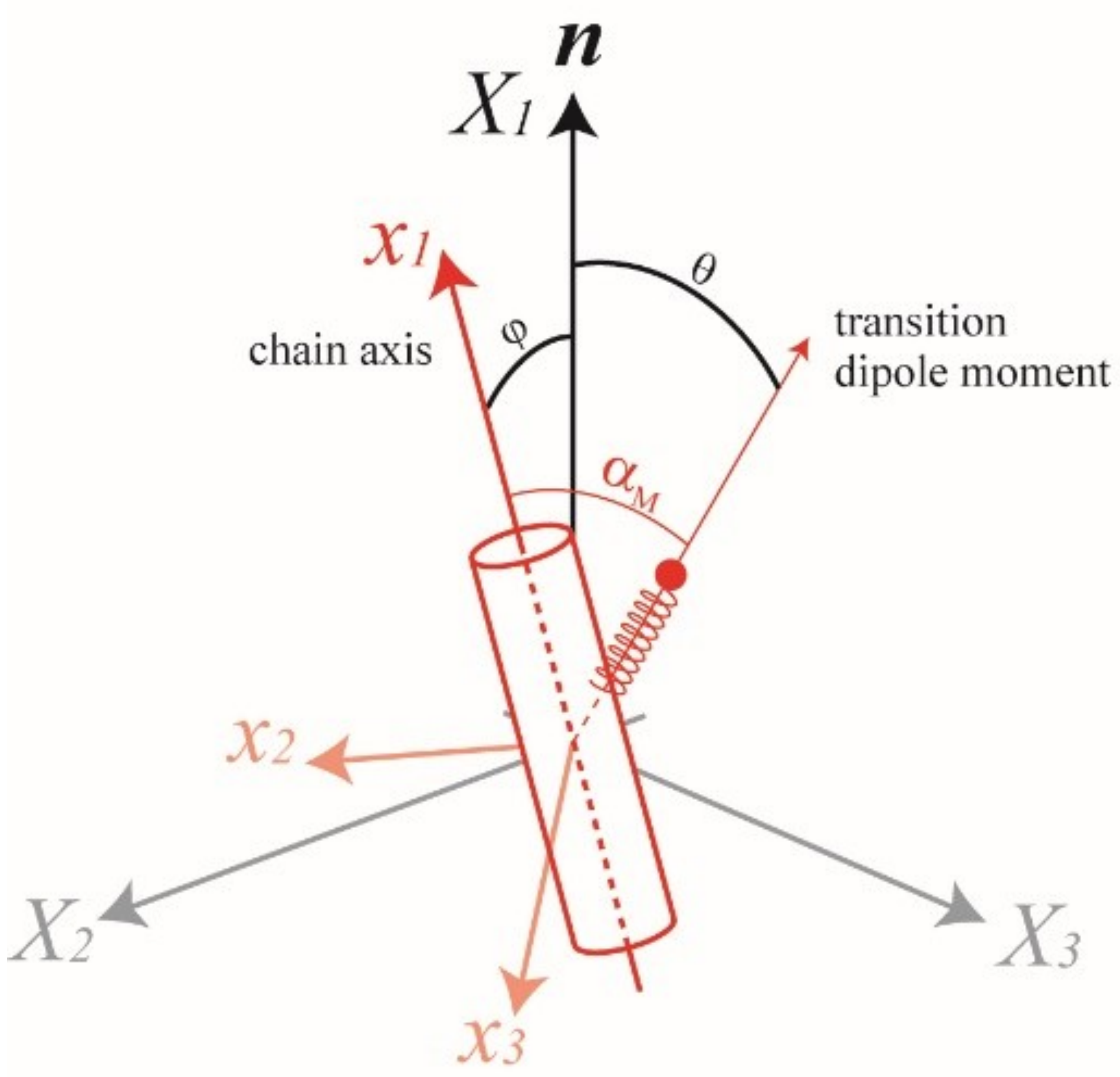}
\caption{Orientation of the transition dipole in respect to a
selected orientation $\mathbf{n}\parallel \mathbf{X}_1$;
$X_1Y_1Z_1$ coordinate system. The orientation of a polymeric
chain is given by direction $\mathbf{x}_1$ in the coordinate
system $x_1y_1z_1$ where $\alpha_m$ defines orientation of the
dipole. The angle $\varphi$ defines tilt between the two
coordinate systems. } \label{f-orient}
\end{center}
\end{figure}
\begin{figure}[htb]
\begin{center}
\includegraphics[width=8.0cm]{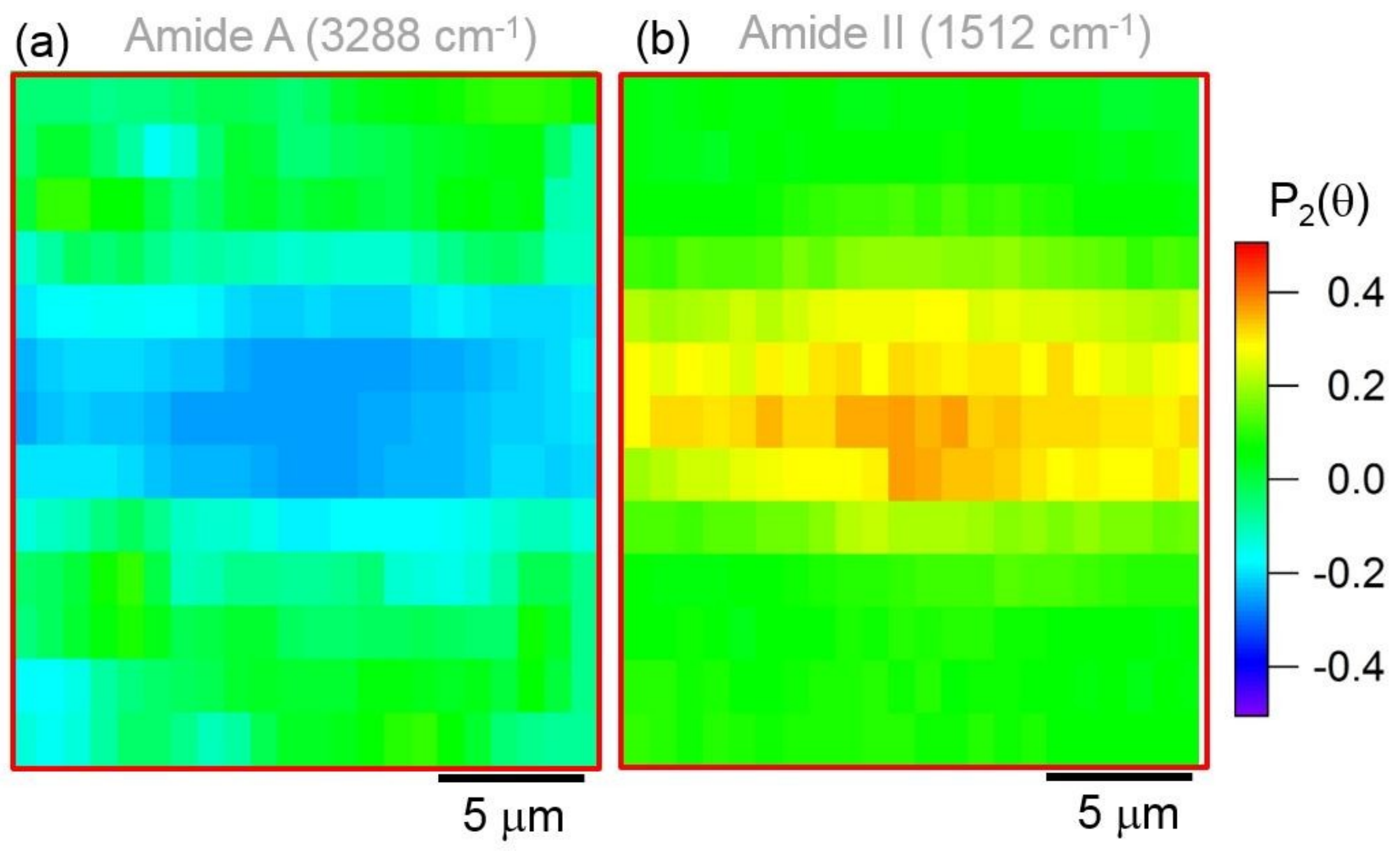}
\caption{The $P_2(\theta)$ order parameter distribution along silk
fiber for different transition dipole moments of Amide A  and
Amide II bands\blue{, which are perpendicular (see, the difference
in color map).}
 } \label{f-sup1}
\end{center}
\end{figure}

Estimation of the polymer chain second order momentum $P_2$ is
carried out considering three angles between the measured
transition dipole moment, $\theta$, the chain axis, $\alpha_m$,
and mutual orientation between polarisation and chain, $\varphi$,
depicted in Fig.~8. The following relations
applies~\cite{Cunningham,Cunningham1}:
\begin{equation}\label{e-m}
P_2 (\theta)=P_2(\alpha_m)\times P_2(\varphi).
\end{equation}
Explicitly,
\begin{equation}\label{em1}
\begin{split}
P_2(\varphi) = \frac{P_2 (\theta)}{P_2(\alpha_m)} &\equiv
\frac{A_1-A_2}{A_1+2A_2}\times\frac{2}{3\langle\cos^2\theta\rangle
- 1} \\
& = \frac{D-1}{D+2}\times\frac{2}{3\langle\cos^2\theta\rangle -
1},
\end{split}
\end{equation}
where $D = A_{0^\circ}/A_{90^\circ} \equiv
A(\varphi_1)/A(\varphi_3)$ is the ratio of the absorbances.

Here, we first estimate the second order parameter $P_2$ of the
measured transition dipole moment which can be directly
compared~\cite{Jarvis} with the value reported in the polarized
Raman scattering~\cite{Rousseau}:
\begin{equation}\label{em2}
P_2 (\theta)= \frac{D-1}{D+2}.
\end{equation}
The strong orientation of C-N, C=O, and N-H present in
$\beta$-sheets can now be clearly revealed by the direct
measurement and polarization analysis from volumes with
cross-sections smaller or comparable with the wavelength not
accessible earlier~\cite{Ling}. The second order parameter,
$P_2(\theta)$ is plotted in Fig.~\ref{f-sup1} for the two
different transition dipole moments showing a prevalent molecular
ordering in the fiber. The molecular alignment in (a) is
perpendicular to the horizontal direction, hence, the value of
$P_2(\theta)<0$ is negative while $P_2(\theta)>0$ (b) for the
transition dipole moment aligned perpendicularly. The epoxy region
surrounding the fiber shows order parameter $P_2(\theta)=0$ as
expected for the random molecular alignment.

For the further evaluation of the second order momentum $P_2$ of
the chain axis (Fig.~8), the value of $\alpha_m$ is required. If
$\alpha_m$ of the Amide A and Amide I is assumed $90^\circ$, then,
$\alpha_m$ of the Amide II can be calculated (Eqn.~6) as
$27.6^\circ$ according to the previous simulation~\cite{Bieri}.
The $P_2(\varphi)$ can be also regarded as an in-plane
distribution.

\begin{table}[h]\caption{The second moment $P_2$ of the orientation function
(Herman's function [-0.5 to 1]) for the Amide bands at wavenumbers
$\lambda$; a negative polariser angle is counted
clockwise.}\label{table}
\begin{tabular}{|c|c|c|c|c|c|c|c|}
 \hline
  Amide&$\lambda$&\multicolumn{2}{|c|}{Absorbance:}&$D$&$P_2(\theta)$&$\alpha_m$&$P_2(\varphi)$\\ \cline{3-4}
       &(cm$^{-1}$)&$0^\circ$&$90^\circ$&$A_{0^\circ}/A_{90^\circ}$& &deg.~\cite{Bieri} & \\
        &             &\blue{$\mathbf{-}$} &\blue{$\mathbf{|}$ }      &        &       &    &\\
  \hline\hline
\blue{A} N-H    &3290  &0.44   &1.12  &0.39\blue{$\pm 0.04$}&\blue{-0.22$\pm 0.026$} &90   &0.51\blue{$\pm 0.05$}   \\
\blue{I} C=O    &1624  &0.33   &1.05  &0.31\blue{$\pm 0.03$}&\blue{-0.29$\pm 0.026$} &90   &0.59\blue{$\pm 0.06$}   \\
\blue{II} C-N   &1510  &1.63   &0.61  &2.67\blue{$\pm 0.3$}&\blue{0.28$\pm 0.042$}  &27.6 &0.53\blue{$\pm 0.06$}   \\
  \hline\end{tabular}
\end{table}

Figure~4 is the spatial distribution of absorbance calculated for
the Amide A, Amide I and Amide II by Eqns.~1-2 with estimation of
$P_2(\theta)$ and $P_2(\varphi)$ summarized in Table~\ref{table}.
The strongest alignment was observed for the C=O bonds which are
participating in the hydrogen bonded $\beta$-sheets
-NH$\cdot\cdot\cdot$O=C- with \blue{$P_2(\theta) = -0.29\pm
0.026$} (uncertainty has been evaluated from three neighboring
pixels along the fiber). This is comparable with $P_2(\theta)= -
0.36$ obtained in Raman scattering from silk
fibers~\cite{Rousseau}; scattering and diffraction anisotropy of
the fiber has an affect onto measurements while flat samples of
the T cross sections were measured in our study.

\bibliographystyle{spiebib}

\end{document}